\documentclass[aip,reprint,onecolumn]{revtex4-1}

\usepackage{amsmath}
\usepackage{amssymb}
\usepackage{graphicx}
\usepackage{color}
\usepackage{float}
\usepackage{sistyle}
\usepackage{subfigure}  
\usepackage{verbatim}   
\usepackage{graphicx}
\usepackage{dcolumn}
\usepackage{bm}
\usepackage[citecolor=blue,filecolor=blue,colorlinks=true,urlcolor=blue]{hyperref}
\usepackage{ulem}
\usepackage{gensymb}
\usepackage{caption}

\begin{document}
	
	\title{Elastic properties assessment in the multiferroic BiFeO$_3$ by pump and probe method}
	
	\author{Pierre Hemme}
	\affiliation{Laboratoire Mat\'eriaux et Ph\'enom$\grave{e}$nes Quantiques (UMR 7162 CNRS), Universit\'e de Paris, Bat. Condorcet, 75205 Paris Cedex 13, France}
	
	\author{Philippe Djemia}
	\affiliation{Laboratoire des Sciences des Procédés et des Matériaux UPR-CNRS 3407, Universit\'e Sorbonne Paris Nord, Alliance Sorbonne Paris Cité, Villetaneuse, 93430, France}
	
	\author{Pauline Rovillain}
    \affiliation{Sorbonne Universit\'e, CNRS UMR 7588, Institut des Nanosciences de Paris, 4 place Jussieu, 75005 Paris, France}
    
	\author{Yann Gallais}
	\author{Alain Sacuto}
	\affiliation{Laboratoire Mat\'eriaux et Ph\'enom$\grave{e}$nes Quantiques (UMR 7162 CNRS), Universit\'e de Paris, Bat. Condorcet, 75205 Paris Cedex 13, France}
	
	\author{Anne Forget}
	\author{Doroth\'ee Colson}
	\affiliation{Service de Physique de l'Etat Condens\'e, DSM/DRECAM/SPEC, CEA Saclay, 91191 Gif-sur-Yvette, France}
	
	\author{Eric Charron}
	\author{Bernard Perrin}
	\author{Laurent Belliard}
	\affiliation{Sorbonne Universit\'e, CNRS UMR 7588, Institut des Nanosciences de Paris, 4 place Jussieu, 75005 Paris, France}
	
	\author{Maximilien Cazayous}\thanks{corresponding author : maximilien.cazayous@u-paris.fr}
	\affiliation{Laboratoire Mat\'eriaux et Ph\'enom$\grave{e}$nes Quantiques (UMR 7162 CNRS), Universit\'e de Paris, Bat. Condorcet, 75205 Paris Cedex 13, France}
	
	\begin{abstract}
		We have performed elasticity measurements in the bulk multiferroic BiFeO$_3$ (BFO) 
		using acoustical pump and probe spectroscopy. The sound velocities of the (quasi)-longitudinal and of the two (quasi)-transverse acoustic waves along three independent directions of the (110) surface have been measured. Moreover, one surface wave and one longitudinal wave propagating perpendicular to the surface have been detected. 
 		Based on initial input values of the six independent C$_{ij}$ elastic constants determined by our density functional theory calculations and our eleven experimental velocities, the numerical resolution of the acoustic equations allows to determine all the C$_{ij}$ elastic constants of BFO. 
		The propagation direction dependence of volume and surface waves phase velocities allows the unambiguously assignment of the waves, hence the polarization of phonons.  
	\end{abstract}
	
	\maketitle
	
	Multiferroics\cite{Eerenstein} that show cross-correlation between electric and magnetic properties are promising materials for devices that transform information from one state into another, such as spin excitations into charge, phonons or photons. There is currently an emerging trend to use these materials in future hybrid computational architectures. In particular the control of spin excitations is highly desirable for information processing in magnonic devices\cite{Sando2013}, just as electro-optics is used to control photonic devices. Their tremendous potential extends to various applications including memories\cite{Scott2007, Martin}, spintronic devices\cite{Dho2006,Allibe2012} or sensors/actuators\cite{Bibes2008}. 
	
	Among them, BiFeO$_3$ (BFO) is one of the very few compound displaying room temperature multiferroicity\cite{Gatalan}. In bulk phase, BiFeO$_3$ is ferroelectric below the Curie temperature $T_C \sim$ 1100 K with a large spontaneous electrical polarization ($\sim$ 100 $\mu$C.cm${^{-2}}$) \cite{Lebeugle2007, Teague1970}.
	The stereochemically active lone-pair electrons present at the Bi sites are responsible of the cationic displacements breaking the inversion symmetry and hence the ferroelectricity \cite{Neaton2005,Ravindran2006}. Also, an antiferromagnetic order appears below the N\'eel temperature $T_N \sim$ 640 K. This magnetic order correspond to a G-type antiferromagnetic modulated by a long-wavelength cycloid of 62 nm \cite{Sosnowska1982}. 
	As a result, Bismuth ferrite has received special attention that led to the discovery of unexpected properties such as conductive domain walls\cite{Seidel2009} or an antiferromagnetic cycloid that can be tuned by strain or electric field\cite{Sando2013, Rovillain2010}. Technological applications have been developed based on this multifunctional material. One can cite BFO-based nanoelectronic devices \cite{Crassous2011} and spintronic \cite{Dho2006,Allibe2012} applications. An additional property that makes BFO not ordinary is its low bandgap unlike usual ferroelectrics. Such characteristic can be used efficiently in photovoltaics \cite{Choi2009,Allibe2010}. It leads also to spectacular THz electromagnetic wave generation and photostriction properties\cite{Talbayev,Kundys, Lejman}.
	The prospects for applications in this area such as optically triggered piezotransducers require a perfect knowledge of the elastic properties of materials. The knowledge of the elastic properties of BFO is desired for the integration of bulk BFO in nano-electronic devices or employing thin films for low cost functional ferroelectric and piezoelectric devices \cite{C8NR05737K}. 
	From an experimental point of view, few works have been devoted to determine its elastic constants. 
	In the work done by Ruello {\it et al.} \cite{Ruello2012}, the sound velocities of one longitudinal and one transverse acoustic phonons along the [110]-pseudocubic direction of BFO have been measured using time resolved optical reflectivity and compared well to previous ab initio calculations. The other directions of propagation have not been studied because additional crystals with different cuts are needed, making it 
	very difficult to determine experimentally the whole set of elastic constants.
	Only inelastic X-ray scattering \cite{Borissenko} succeeded to deduce the elastic constants set of BFO via the resolution of the Christoffel equations using the sound velocity measured along high-symmetry directions.  For the majority of constants, the resolution is only based on two measured points in the linear part of the acoustic phonon dispersion curve which leads to an important error bar. 
	
	In this letter, we implement acoustical pump-probe experiment at the nanosecond time-scale on BiFeO$_3$ single crystal. 
	We measure the sound velocities of both longitudinal and transverse acoustic modes traveling along different directions in the (110)-rhombohedric plane. 
	From a set of initial elastic constants calculated by the density functional theory, 
	the resolution of the Christoffel equation gives access to first theoretical velocities. The minimization of the difference between the experimental and theoretical velocities allows to determine all six independent C$_{ij}$ elastic constants.
	 The comparison of the directional dependency of experimental and theoretical sound velocities ($V$) enables identification of the longitudinal (L), 
	fast (S1) and slow (S2) transversal acoustic modes and the generalized Rayleigh surface wave (GRSW).
	
	
The BiFeO$_3$ single crystals studied were grown in air using a Bi$_2$O$_3$-Fe$_2$O$_3$ flux technique and present a single ferroelectric and magnetic domain state \cite{Lebeugle2007}. The samples have a millimeter size with a large surface plane corresponding to the $(010)_{pc}$ or to the $(110)_{rh}$ face in the pseudo-cubic or rhombohedric representation, respectively. 
Aluminium thin film of 75 nm has been evaporated on the BFO surface. The pump wavelength is below the gap of BiFeO3 then Al thin film acts as a transducer when illuminated by the pump.	
It also enhances the wave observation by increasing the reflectivity signal of the probe. The optical penetration depth in BFO is quite large in the optical wavelength domain used in our experiment (BFO is nearly transparent) making the detection of surface displacements more complicated without Al and adding strong mixed Brillouin signals (see supplemental). Al allows to detect essentially the surface signal and thus to measure the surface displacements owing to the interferometric detection with a limited contribution of the photoelastic coupling from the bulk. The thickness of Al is very small compared to the wavelengths of the acoustic waves involved in our experiment.  If the Al thin film has an effect, it could only increase moderately the uncertainties of the determined elastic coefficients. Notice that, BFO single crystals have a small piezoelectric response \cite{Lebeugle2007, Rovillain2009} compared to excellent piezoelectric compounds.\cite{Guennou} This property was therefore not considered in our analysis.
	
\begin{figure}[h!]
		\begin{center}
			\includegraphics[width=6.4cm]{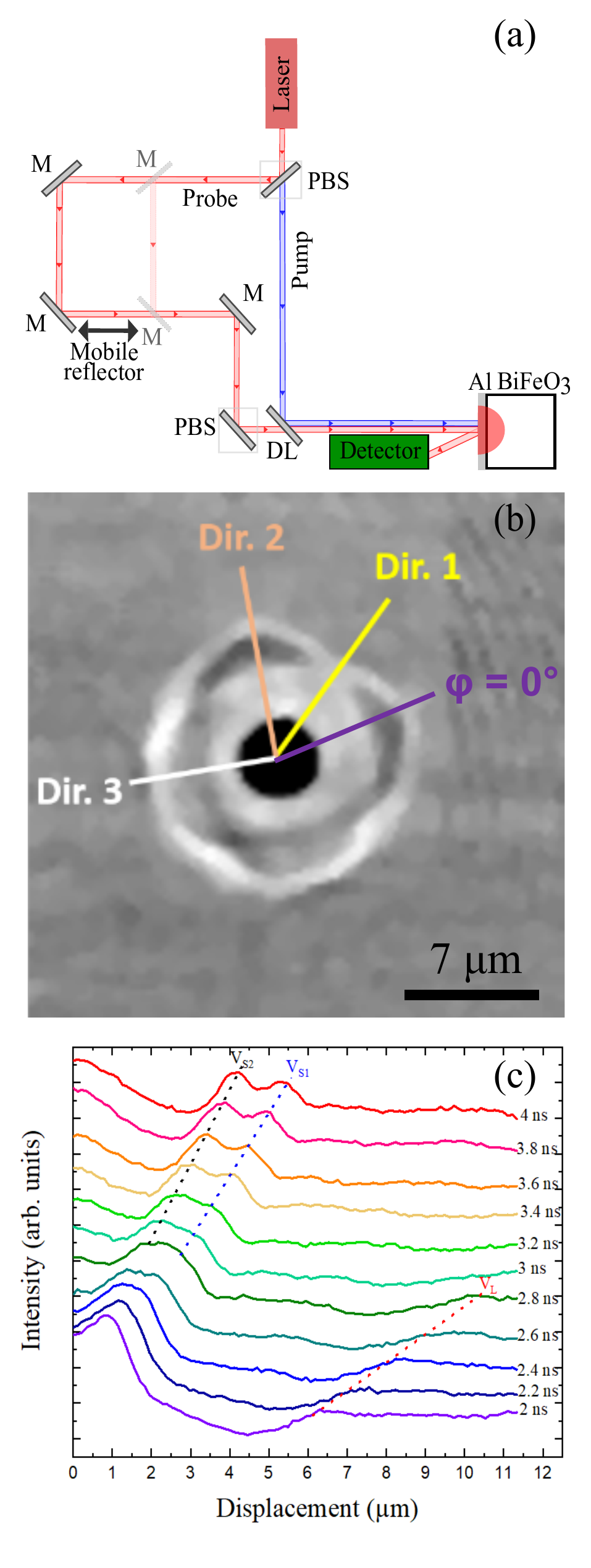}
			\caption{(a) Sketch of the experimental setup with first the pump focused on BFO capped with an Al thin film, second the probe path (PBS polarized beam splitters, M are mirrors) and the mobile reflector for the time delay, third the detection path. 
			(b) Scan of the BFO $(010)_{pc}$ surface at fixed pump-probe delay of 3.6 ns showing the propagating acoustic waves emerging from the epicenter. (c) Relative phase change of the electromagnetic field of the probe beam as a function of the displacement along "direction 1" (distance in $\mu$m measured from the center of the black area in Fig. 1(a)) and at different probe time delay.}
			\label{Fig1}
		\end{center}
\end{figure}

	Acoustical pump-probe technique is an optical method which generates short acoustic pulses which can be detected by a time-delayed probe beam. Owing to the large frequency bandwith of the acoustic pulse, it is thus possible to study materials at a submicron scale. 
    This time resolved measurements allows to measure sound velocity with an accuracy typically better than 5$\%$, and thus the related elasticity properties.
	
	For our pump‐probe experiment\cite{Xu, Amziane, Belliard} as shown in Fig.\ref{Fig1}(a), a mode-locked Ti:sapphire (MAI TAI Spectra) laser source operating at 800 nm was used to thermally excite and optically detect the propagating acoustic modes in BFO. The pulse duration is 200 fs and the pulse repetition rate is 80 MHz. 
	In order to get some above band gap excitation, the BFO sample was excited by light obtained from second harmonic
	generation (SHG) with wavelength 400 nm performed by doubling the pump frequency with a nonlinear crystal (BBO). The power of the two beams is fixed at 300 $\mu$W. 
    Acoustic measurements were performed using a standard stabilized Michelson interferometer sensitive to the perpendicular surface displacement. A 12 ns maximum pump-probe time delay is achieved using a mobile reflector mounted on a translation stage. Both the pump and probe beams are focused with a microscope objective with a numerical aperture of 0.9, fixed on a piezoelectric stage. A tilting system of the probe allows the mapping of acoustic waves propagating in any direction of the  (010)$_{pc}$ plane, thus providing information on the elastic properties of BiFeO$_3$ single crystals.  
	
	Figure \ref{Fig1}(b) shows the propagation of the acoustic waves on the $(010)_{pc}$ BFO surface. The signal is recorded by scanning the surface of the sample (28$\times$28 $\mu$m$^2$) at fixed pump-probe delay. It corresponds to a snapshot taken after 3.6 ns from the coincidence (temporal overlay of the pump and probe). The black area in the center of the image is a huge stationary photothermal component. The contrast in intensity corresponds to maxima and minima in the amplitude of the wavefront. We have performed scan lines along three independent directions giving the profile of the acoustic waves at different times of probe arrival after the coincidence, from 2 ns to 4 ns by step of 200 ps. Figure \ref{Fig1}(c) presents the relative change of the probe reflectivity along the scanned "direction 1" as a function of the probe position and for several probe time delays. One can observe in this figure three peaks whose positions evolve as a function of the probe delay. These peaks correspond to wave amplitude maxima and three distinct propagation speeds. Remember that our measurements are performed on a $(010)_{pc}$ face of the pseudocubic-oriented crystal allowing the observation of one longitudinal acoustic (LA) and two non-degenerate transverse acoustic (TA) phonons. The exact attribution and description of these peaks is done later in the article. 
	
	In order to attribute these wave propagation speeds to the correct acoustic modes, we have implemented several calculation steps to optimize the elastic constants by minimizing the difference between the experimental and theoretical velocities.
	Knowing the C$_{ij}$ elastic constants and mass density (we used $\rho=$8330 kg/m$^{3}$) of the material, allows to predict sound velocities by solving the Christoffel equation. According to the BFO  space-group symmetry (R3c), the elastic constant tensor C$_{ij}$ has seven independent constants \cite{Born}, namely, C$_{11}$, C$_{13}$, C$_{14}$, C$_{25}$, C$_{33}$, C$_{44}$, C$_{66}$=$\frac{C_{11}-C_{12}}{2}$, sufficient to simulate all the elastic properties of the material.
	The initial input values C$_{ij}$ were obtained by density functional theory (DFT) calculations.\cite{Dreizler}
	The VASP package\cite{Hafner2008} and the electron-ion interaction, described via the projector augmented wave method, are used to perform first-principles calculations.\cite{Kresse}
	The generalized gradient approximation (GGA) is implemented to depict the exchange correction functional, specifically we used both the Perdew–Burke–Ernzerhof (PBE) and the one revised for solids (PBEsol) \cite{Perdew} to vary the equilibrium atomic volume at which the elastic properties are determined. This is a spin-polarized magnetic calculation, with electronic iterations convergence of 0.001 meV using the Normal (blocked Davidson) algorithm, plane wave cutoff energy of 520 eV and reciprocal space projection operators. The Brillouin-zone {\it k}-mesh is forced to be centered on the gamma point and corresponds to actual {\it k}-spacings of 15$\times$15$\times$15. The stress-strain method is employed with applied strain amplitudes $\pm$ 0.001, $\pm$ 0.0025 and $\pm$ 0.005, and elastic constants are determined on the relaxed structures by using the tetrahedron method incorporating Bl\"{o}chl corrections.\cite{Blochl1994} 

Thus, the initial theoretical velocities can be calculated from DFT C$_{ij}$'s (see Tab.1). To compare theoretical and experimental velocities, a $\chi^2$ test is used. The least squares method allows to converge towards optimized elastic constants which reproduced well our experimental findings while explicit dependencies are obtained from the analytical formula shown in the supplementary materials.

		\begin{table} [h]    
		\begin{ruledtabular}
			\begin{tabular}{c|c|c|c|c|c}
				{C$_{ij}$} & {DFT$^*$} & {DFT$^{\cite{Shang}}$} & {Exp$^{\cite{Ruello2012}}$} & {Exp$^{\cite{Borissenko}}$} & {Final set}\\
				\hline
				C$_{11}$ = C$_{22}$						  & 177-220 & 203-222 & 205-255 & 249 & 215   \\ 
				C$_{33}$           						  & 116-140  & 129-150 &     & 160 & 180   \\ 
				C$_{44}$ = C$_{55}$						  & 48-63   &  31-49 &     &  44 &  40   \\
				C$_{66}$           						  & 43-51   &  43-56 &     &  49 &  60   \\ 
				C$_{12}$ = C$_{11}$-2C$_{66}$	          & 118-92  & 110-117 &     & 151 & 95   \\
				C$_{13}$           						  & 40-68   & 50-50  &     &  75 &  50   \\ 
				C$_{14}$           						  & 6-12    & 16-23  &     &  9  &  27   \\
				C$_{25}$=C$_{46}$=-C$_{15}$           						  & $\approx$ 0    &   &     &    & 
				 \\ 
				V$_{0}$           						  & 11.52-12.37    & 12.66-12.74  &      &    &  
				 \\ 
				V$_{0}$ (Exp.)           						  &     &   & 12.46     &    & 
				\\
			\end{tabular}
		\end{ruledtabular}
		\caption{C$_{ij}$ elastic constants determined in this work by DFT (*) with GGA-PBE and GGA-PBEsol exchange-correlation functional and compared to values already calculated by DFT\cite{Shang} with GGA-PBE$+$U (U = 0 and 6 eV) simplified approach of
Dudarev {\it et al.} \cite{Dudarev1998} or determined by time domain Brillouin spectroscopy  \cite{Ruello2012} and by inelastic X-ray scattering \cite{Borissenko}. Elastic constants values are given in GPa and the equilibrium atomic volume V$_0$ in {\AA}$^3$/atom.}
		\label{Tab1}
	\end{table}

	Table \ref{Tab1} summarizes the results of this procedure using the velocities obtained in the three directions and the comparison with previous theoretical and experimental values. The final values of the elastic constants deviate by less than 10$\%$ from the previously calculated or experimental values except for C$_{44}$, C$_{33}$ and C$_{14}$ for which the deviation is about 20$\%$ and 60$\%$, respectively. Compared to previous work, a greater number of velocities along several directions are experimentally determined here. In particular, prior knowledge of BFO refractive index is avoided, in contrary to the time-domain Brillouin scattering (TDBS) conducted by Ruello {\it et al.} along the [010]$_{rh}$ direction 
	(see section II.B of supplementary materials for our TDBS measurements of V$_L$ and V$_T$ along [110]$_{rh}$). Of course, more experimental velocities are measured, more the resolution of the inverse problem gives precise elastic constants. 
	
		\begin{figure}[h!]
		\begin{center}
			\includegraphics[width=8.5cm]{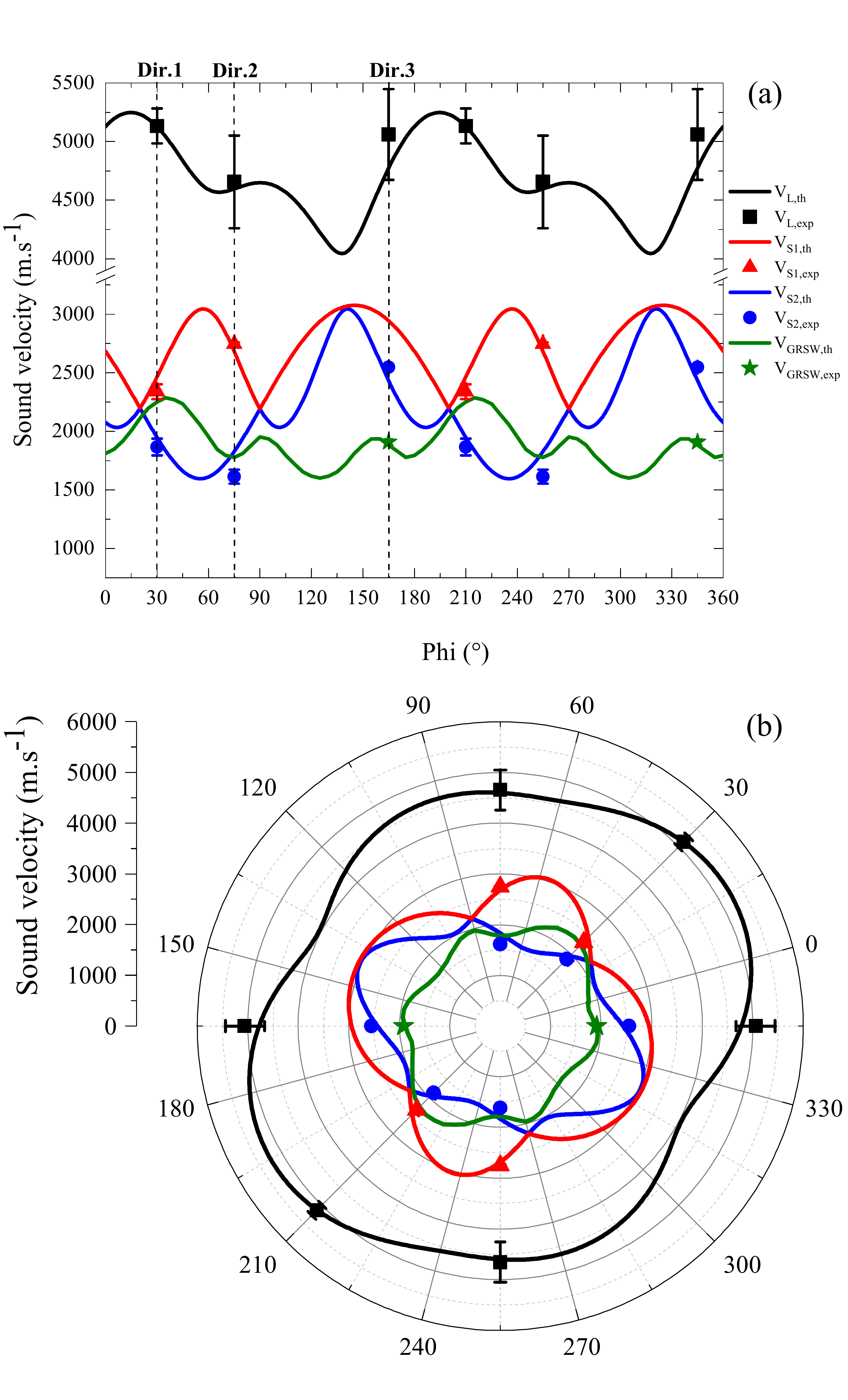}
			\caption{(a) Experimental (symbols) and calculated (lines) sound velocities of bulk (V$_L$, V$_{S1}$, V$_{S2}$) and surface acoustic waves (V$_{GRSW}$) as a function of the $\phi$ angle in respect to the ($\bf{a-b}$) direction. Vertical lines represent the three scanning directions. (b) Polar representation of the data in the (110)$_{rh}$ plane.}
			\label{Fig2}
		\end{center}
	\end{figure}
	
	We now detail the calculations of the sound velocities of bulk waves using analytical expressions (see the supplementary materials) and we determine the acoustic modes involved, which will allow a direct comparison with the experimental pattern in Fig. \ref{Fig1}(a).
	The Christoffel equations can be easily derived. For the $(110)_{rh}$ plane, the reference axis-frame we used for the elastic tensor, is ($\bf{a-b}$, $\bf{c}$, -$\bf{(a+b)}$) with ($\bf{a}$, $\bf{b}$, $\bf{c}$) the rhombohedric lattice. The direction of the propagative wave is defined by the angle $\phi$ in respect to the ($\bf{a-b}$) direction. The particular direction $\phi$=0$^{\circ}$ was afterwards reported in Fig. \ref{Fig1}(b). In case of surface waves, we used a numerical simulation program that solves the acoustic problem of a semi-infinite anisotropic elastic medium. \cite{djemia2004, Djemia1998} 
	
	Figure \ref{Fig2}(a) presents the calculated and experimental velocities associated to the three bulk waves and the surface wave as a function of the $\phi$ angle. The experimental points belong only to the direction 1, 2 and 3. From this single figure, the non-trivial evolution of the slowest sound velocities V$_{S1}$, V$_{S2}$ and V$_{GRSW}$ obviously cannot be immediately assessed, while the fastest longitudinal wave is only visible at lowest probe delay time below $\sim$2.8 ns (see Fig. \ref{Fig1}(c)). 
	The wave with the fastest velocity between 4000 and 5500 m.s$^{-1}$ corresponds to the longitudinal acoustic phonon. 
	The other slower waves with velocities between 1250 and 3000 m.s$^{-1}$ are associated to the two (quasi)-transverse acoustic phonons and a generalized Rayleigh surface wave. 
	The two shear waves are artificially but conveniently differentiated by their speed, the fast transverse mode always (S1) having a faster speed than the slower transverse mode (S2) and the surface wave (GRSW). Graphically in the Fig. \ref{Fig1}(b), only two wave fronts are clearly visible. We can see some crossing between the slow waves such as for the "direction 1" or repulsing like in  "direction 2" and "direction 3".
	Thanks to our analytical and numerical resolution of the acoustic eigenvalues problem, it is possible to track unambiguously each wave by identifying simultaneously their sound velocities (eigen values) and their necessarily different polarizations (eigen vectors) as a function of $\phi$. Such identifications are provided in case of the bulk waves and examplified for the surface wave, in the supplementary materials.
	
	One velocity along direction 3 (star in \ref{Fig2}(a)) is below from the calculated bulk shear velocities, and can be identify to a true surface wave. To find an explanation for this apparent discrepancy, we have calculated with a home-made program, \cite{djemia2004,Djemia1998} the generalized Rayleigh surface wave (GRSW) for a semi-infinite anisotropic elastic medium. The GRSWs can propagate only along some directions in anisotropic crystals with a majority sagittal polarization and a sound velocity lower than any of the two bulk shear waves for a true surface wave or above in case of a pseudo-surface mode.
	One can see the good agreement between the experimental velocity and the calculated GRSW one, even if this comparison is only based on this unique analyzed experimental point. The generalized Rayleigh surface wave is expected to be observed in the vicinity of "direction 3", $\approx$165$^{\circ}$ away from the ($\bf{a-b}$) direction in the $(110)_{rh}$ plan. 
	
	Figure \ref{Fig2}(b) shows the wave front curves within a surface cut in the $(110)_{rh}$ plane that can be compared to Fig. \ref{Fig1}(b). As-mentioned before, the longitudinal acoustic phonon is too fast and it is no longer present in the image of Fig. \ref{Fig1}(b) while the other slower waves are recorded. In Fig. \ref{Fig2}(b) a flower shape can be observed, reproducing the experimental pattern with four petals in the vicinity of the same directions. Notice that, we have not included in the simulations the aluminium thin film. This addition should explain the observation of all polarizations (longitudinal, horizontal transverse and vertical transverse). In this system, mode conversion should exist at the Al/BFO interface. All these volume or surface modes have thus a small vertical component at the free surface of Al that plausibly allows their observation while it may remain a photoelastic coupling limited the near surface, too.

	To conclude, we determined experimentally all six non-zero independent C$_{ij}$ elastic constants of BiFeO$_3$ by acoustical pump-probe method at the nanosecond time-scale. We considered a minimum of 3 independent scanning directions, enabling measurement for each, of three sound velocities of bulk and surface waves. We have highlighted the directional dependence of the sound velocities of the longitudinal and two transverse bulk acoustic phonons as well as the generalized Rayleigh surface wave. 
	The experimental values of the elastic constants are highly desired for BFO integration in micro- and nanoelectronic devices, as for stresses evaluation in thin films. 
	
\subsection*{SUPPLEMENTARY MATERIAL}
See supplementary material for the equations connecting the sound velocities of bulk waves to the C$_{ij}$ elastic constants, time-domain Brillouin scattering experiments and the assessment of the generalized Rayleigh surface wave sound velocities.

\subsection*{ACKNOWLEDGMENTS}
The authors acknowledge J. Rastikian and S. Suffit for the evaporation of aluminium thin film on the BFO surface at the cleanroom of Université de Paris. 

\subsection*{AIP PUBLISHING DATA SHARING POLICY}
The data that supports the findings of this study are available within the article [and its supplementary material].


\newpage
\newpage
	
	\section{{Bulk waves sound velocities by solving the Christoffel equations}}
	
	In this supplementary material, we provide first the equations connecting the sound velocities of bulk waves to the C$_{ij}$ elastic constants.
	
	In the rhombohedral representation ({\bf a}, {\bf b}, {\bf c}) of BiFeO$_3$ with angles $\alpha=\beta=90$ and $\gamma=120^{\circ}$, the symmetry of the elastic constant tensor is trigonal with six independent elastic constants, neglecting the others C$_{25}$=C$_{46}$=-C$_{15}$$\approx 0$:
	
	\begin{equation}
	C_{{\bf a}, {\bf b}, {\bf c}} = 
	\begin{pmatrix}
	C_{11} & C_{12} & C_{13} & C_{14} & 0 & 0 \\
	C_{12} & C_{11} & C_{13} & -C_{14} & 0 & 0 \\
	C_{13} & C_{13} & C_{33} & 0 & 0 & 0 \\
	C_{14} & -C_{14} & 0 & C_{44} & 0 & 0 \\
	0 & 0 & 0 & 0 & C_{44} & C_{14} \\
	0 & 0 & 0 & 0 & C_{14} & C_{66}=\frac{C_{11}-C_{12}}{2} \\
	\end{pmatrix}
	\end{equation}
	
	In the experimental (110) plane, the new orthogonal representation is ({\bf a-b}, {\bf c}, -({\bf a+b}). This is obtained by the rotation of -30$\degree$/{\bf c} and -90$\degree$/({\bf a-b}). The tensor of the elastic constants is now written :
	
	\begin{equation}
	C_{{\bf a-b}, {\bf c}, {\bf -(a+b)}} = 
	\begin{pmatrix}
	C_{11} & C_{13} & C_{12} & 0 & 0 & C_{14} \\
	C_{13} & C_{33} & C_{13} & 0 & 0 & 0 \\
	C_{12} & C_{13} & C_{11} & 0 & 0 & -C_{14} \\
	0 & 0 & 0 & C_{44} & -C_{14} & 0 \\
	0 & 0 & 0 & -C_{14} & C_{66}=\frac{C_{11}-C_{12}}{2} & 0 \\
	C_{14} & 0 & -C_{14} & 0 & 0 & C_{44} \\
	\end{pmatrix}
	\end{equation}
	
	The propagation of a bulk acoustic wave is described by Christoffel's equation :
	
	\begin{equation}
	\rho\frac{\partial^2\vec{u}}{\partial t^2}=\vec{\nabla}\sigma
	\end{equation}
	
	with $\rho$ the volumic mass, $\vec{u}=\vec{u}_0e^{i(\vec{k}\vec{r}-\omega t)}$ the displacement vector where $\vec{k}$ is the wave vector, $\sigma_i$ is the stress tensor equal to $C_{ij}\epsilon_j$, $\epsilon$ being the deformation tensor. The phase speed is defined by : V=$\frac{\omega}{k}$.\\
	All following analytical results were checked by a fully numerical resolution using the program "Christoffel" provided in Refs. \onlinecite{JAEKEN2016a,Jaeken2016b}.
	
	\subsection{Propagation in the (110) plan}
	\begin{figure}[h]
		\begin{center}
			\includegraphics[width=8.5cm]{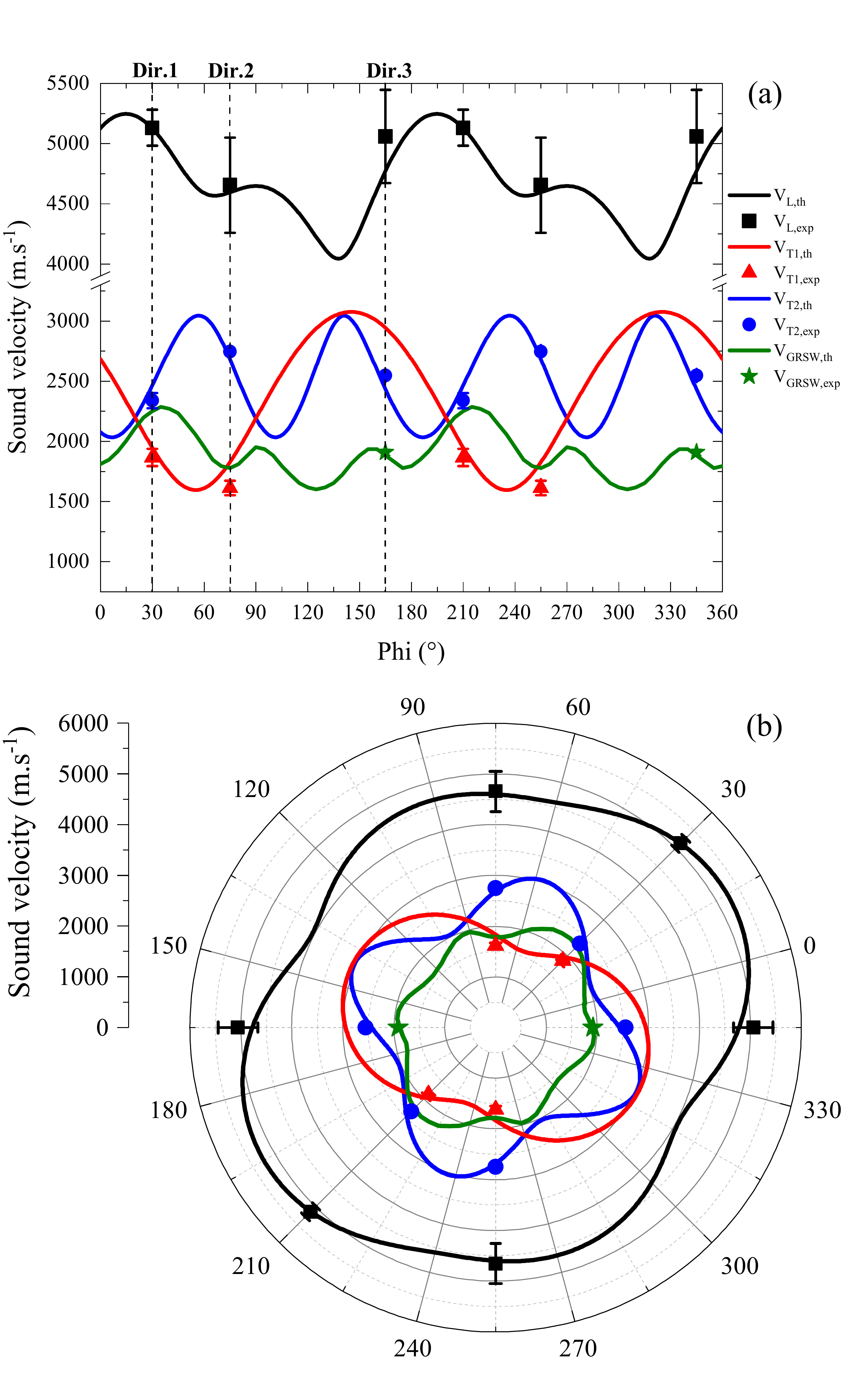}
			\caption{(a) Experimental (symbols) and calculated (lines) sound velocities of bulk (V$_L$, V$_{T1}$, V$_{T2}$) and surface acoustic waves (V$_{GRSW}$) as a function of the $\phi$ angle in respect to the ($\bf{a-b}$) direction. Vertical lines represent the three scanning directions. (b) Polar representation of the data in the (110)$_{rh}$ plane.}
			\label{Fig2_supp}
		\end{center}
	\end{figure}
	
	The wave vector is defined by  $(kcos(\phi),ksin(\phi),0)$ with respect to the {\bf(a-b)} direction ($\phi=0$).  
	The solutions in agreement with Ref. \onlinecite{Pace1971} are:
	
	1. a quasi-longitudinal mode polarized along (U$_1$,u$_2$)  with speed equal to V$_{L}$ = V$_+$,\\
	
	2. a quasi-transverse mode polarized along (u$_1$,U$_2$)  with speed equal to V$_{T1}$ = V$_-$:
	\begin{equation}
	\rho {V^2_{+/-}}=\frac{b \pm \sqrt{b^2-4c}}{2} 
	\end{equation}
	\\
	with 
	\begin{equation}
	b=(C_{44}+C_{11})cos^2(\phi) +(C_{44}+C_{33})sin^2(\phi) +2C_{14} cos(\phi)sin(\phi)
	\end{equation}
	\begin{equation}
	c=-[C_{14} cos^2(\phi) +(C_{44}+C_{13})cos(\phi) sin(\phi)]^2+(C_{11}cos^2(\phi)+C_{44}sin^2(\phi)+2C_{14} cos(\phi)sin(\phi))(C_{44} cos^2(\phi)+C_{33}sin^2(\phi))
	\end{equation}
	3. A pure transverse mode vertically polarized along (U$_3$) with a velocity V$_{T2}$:
	\begin{equation}
	\rho {V^2_{T2}}=\frac{(C_{11}-C_{12})}{2}cos^2(\phi)+C_{44}sin^2(\phi)-2C_{14} cos(\phi)sin(\phi))
	\end{equation}
	
	The directional dependence of those three sound velocities is displayed in Fig \ref{Fig2_supp}, together with our picosecond ultrasound experimental data in the Letter. 

	\subsection{Propagation along the [110] direction and time-domain Brillouin scattering experiments}
	
	The wave vector along the [110] direction is $(0,0,k)$. The solutions in agreement with Ref. \cite{Pace1971} are:
	
	1. a quasi-transverse (fast) mode polarized along (U$_1$,u$_2$)  with speed equal to V$_{T2}$ = V$_+$,\\
	
	2. a quasi-transverse (slow) mode polarized along (u$_1$,U$_2$)  with speed equal to V$_{T1}$ = V$_-$:
	\begin{equation}
	\rho {V^2_{+/-}}=\frac{C_{44}+C_{66} \pm \sqrt{(C_{44}+C_{66})^2+4(C_{14}^2-C_{44}C_{66})}}{2}
	\end{equation}
	
	3. A longitudinal mode polarized according to (U$_3$) with a velocity V$_L$ :
	\begin{equation}
	\rho {V^2_L}=C_{11}
	\end{equation}

	We have performed time-domain Brillouin scattering on BiFeO$_3$ single crystal along the [110] direction of the rhombohedral representation. This experiment is very similar to the one conducted by Ruello {\it et al.} in Ref. \onlinecite{Ruello2012} but along the [010] direction.
	\begin{figure}[h]
		\begin{center}
			\includegraphics[width=15cm]{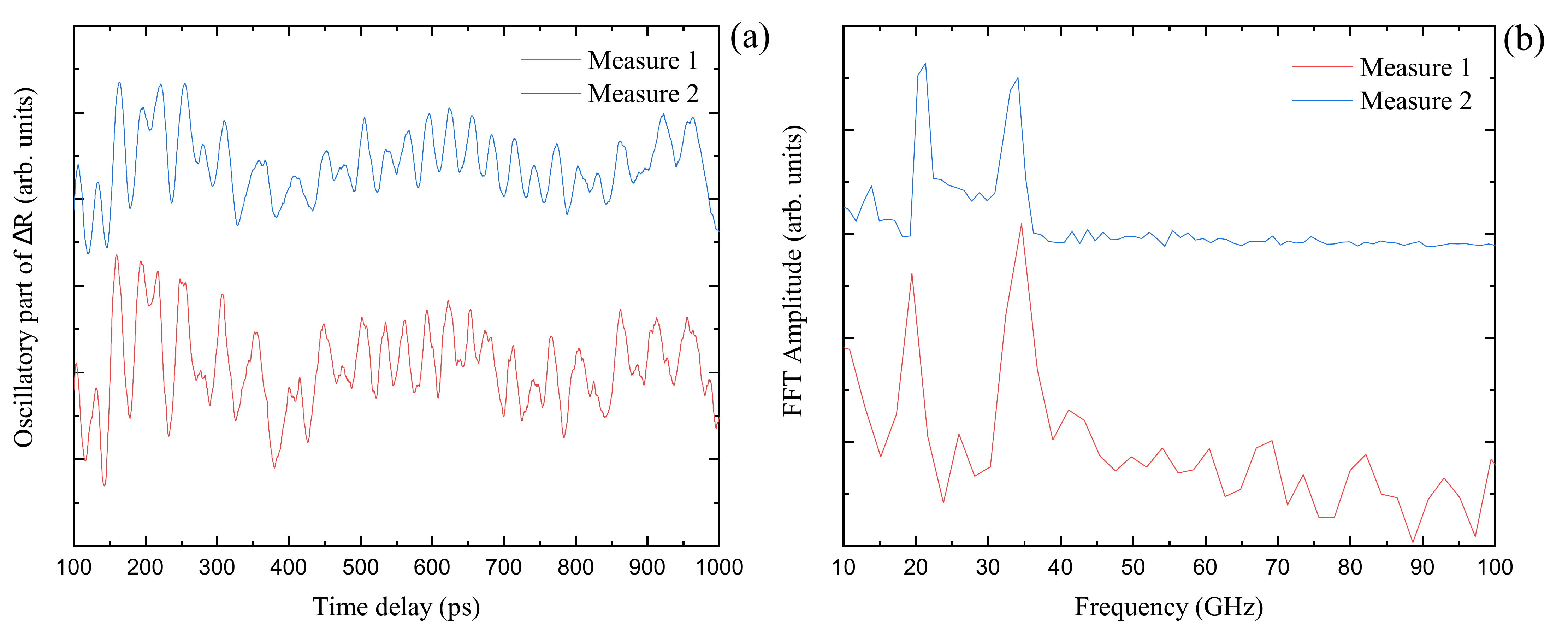}
			\caption{(a) Brillouin oscillations detected along the [110] direction for two repeated experimental runs (measures 1 and 2). (b) Fast Fourier Transform of the associated oscillations, revealing two acoustic modes at $\approx 20$ and $\approx 33$ GHz.}
			\label{OS}
		\end{center}
	\end{figure}

	
	Figure \ref{OS} shows the Fast Fourier transform of the oscillations observed in the time resolved optical reflectivity. Two doublets are revealed at 20 GHz and at 33 GHz. They correspond to the  fast shear (T2) and longitudinal (L) acoustic modes, respectively.
	
	In the back-scattering configuration used in our  experiment, the momentum conservation is obtained for $\vec{q}_{ac}$=2$\vec{k}_{op}$ with $\vec{q}_{ac}$ the momentum of acoustic phonon and $\vec{k}_{op}$ the momentum of photon inside BFO. $k_{op}$=2$\pi n/\lambda$ where $\lambda$ is the wavelength of the laser probe and $n$ is the averaged refractive index of BFO at the wavelength $\lambda$. The phonon dispersion law is assumed to be written as $2\pi f_{ac}$=V$q_{ac}$ where V is the sound velocity in BFO associated to the Brillouin frequency $f_{ac}$. We can thus deduce the velocity of the fast shear wave (V$_{T2}$=V$_+$) and longitudinal wave (V$_L$) from Fig. \ref{OS} using $\lambda$=797.5 nm and $n$=2.6-2.9: \cite{Kumar2008,Gu2009}
	
	\begin{equation*}
	V_{T2} = 2953-2648 \:  m.s^{-1}\: \text{and} \:
	V_L = 5061-4538\: m.s^{-1}
	\end{equation*}
	
	If we consider the lower index $n$=2.6, we find the same velocity V$_L$ (~5080 m.s$^{-1}$) measured along [1$\bar{1}$0] in our Letter and thus the same C$_{11}$. It is the same for V$_{T2}$ (~3075 m.s$^{-1}$) that leads to an additional constraint on C$_{44}$, C$_{66}$ and C$_{14}$ which was found to be of importance in the fitting procedure. Contrary to our approach, the time-domain Brillouin scattering allows only to measure few directions around the normal direction and necessitates additional crystal with different orientations for further scrutinizing elastic properties. 
	
	\section{Assessment of the generalized Rayleigh surface wave sound velocities}
	In case of surface waves, we search for partial displacement vector $\vec{u}=\vec{u}_0e^{ik_zz}e^{i(\vec{Q}_{//}\vec{r}_{//}-\omega t)}$, with the addition of a free-stresses boundary condition at the free surface  ($z$=0): $\sigma_{jz}=0$ (for $j=x,y,z$), from which the three unknown $k_z$ complex parameters are determined. ${\vec{Q}}_{//}$ and $\omega$ are, respectively, the in-plane wave vector (spanning any directions from ({\bf a-b}) to ({\bf c}) and the circular frequency of the surface acoustic wave, from which its sound velocity $V=\frac{\omega}{Q_{//}}$ is defined.
	
	\begin{figure}[h]
		\begin{center}
			\includegraphics[width=9cm]{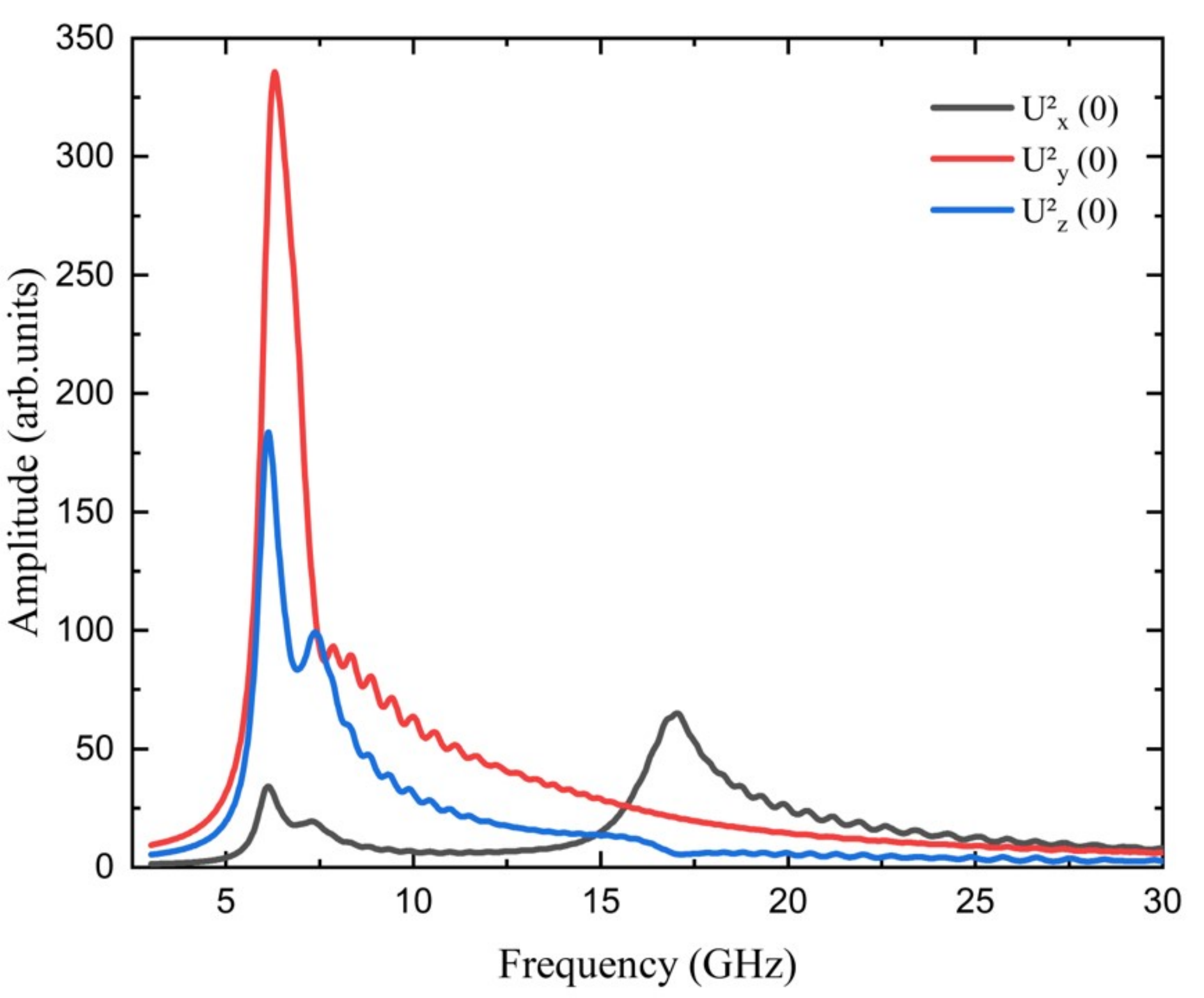}
			\caption{Spectral density for the angle Phi = 165$^{\circ}$.}
			\label{S3}
		\end{center}
	\end{figure}
	
	Employing a home-made program,\cite{Djemia2004,Djemia2001,Djemia1998} we performed numerical calculations of spectral densities   $<u^2_j(z=0)>_{{\vec{Q}}_{//}, \omega}$ related to the surface waves displacement components at the free surface with a fixed Q$_{//}$ wave vector modulus. It is illustrated in Fig \ref{S3}, peaks maximum providing the surface wave frequency.



\subsection*{REFERENCES}
	\begin{thebibliography}{1}
	\bibitem{Eerenstein} W. Eerenstein, N. D. Mathur, J. F. Scott, Nature {\bf 442}, 759 (2006).
		
	\bibitem{Sando2013} D. Sando, A. Agbelele, D. Rahmedov, J. Liu, P. Rovillain, C. Toulouse, I. C. Infante, A. P. Pyatakov, S. Fusil, E. Jacquet, C. Carrétéro, C. Deranlot, S. Lisenkov, D. Wang, J.-M. M. Le Breton, M. Cazayous, A. Sacuto, J. Juraszek, A. K. Zvezdin, L. Bellaiche, B. Dkhil, A. Barthélémy, and M. Bibes, Nat. Mater. {\bf 12}, 641 (2013).
		
	\bibitem{Scott2007} J. F. Scott, Nat. Mater. {\bf 6}, 256 (2007).
		
	\bibitem{Martin} L. W. Martin, Y-H Chun, and R. Ramesh, Emerging Non-Volatile Memories in S. Hong , O.Auciello, and D. Wouters (eds) Emerging Non-Volatile Memories. Springer, Boston, MA (2014).
		
	\bibitem{Dho2006} J. Dho, X. Qi, H. Kim, J. L. MacManus-Driscoll, and M. G. Blamire, Adv. Mat. {\bf 18}, 1445 (2006).
		
	\bibitem{Allibe2012}  J. Allibe, S. Fusil, K. Bouzehouane, C. Daumont, D. Sando, E. Jacquet, C. Deranlot, M. Bibes, and A. Barth\'el\'emy, Nano Lett. {\bf 12}, 1141 (2012).
		
	\bibitem{Bibes2008} M. Bibes, and A. Barthelemy, Nat. Mater. {\bf 7}, 425 (2008).
		
	\bibitem{Gatalan} G. Catalan and J. F. Scott, Adv. Mater. {\bf 21}, 2463 (2009).
	
	\bibitem{Lebeugle2007}  D. Lebeugle, D. Colson, A. Forget, M. Viret, P. Bonville, J. Marucco, and S. Fusil, Phys. Rev. B  {\bf 76}, 024116 (2007).
		
	\bibitem{Teague1970}  J. R. Teague, R. Gerson,  and W. J. James,  Solid State Commun. {\bf 8}, 1073 (1970).
		
	\bibitem{Neaton2005}  J. Neaton, C. Ederer, U. Waghmare, N. Spaldin, and K. Rabe,  Phys. Rev. B {\bf 71}, 014113 (2005).
		
	\bibitem{Ravindran2006}  P. Ravindran, R. Vidya, A. Kjekshus, H. Fjellv\.{a}g, and O. Eriksson, Phys. Rev. B {\bf 74}, 224412 (2006).
	
	\bibitem{Sosnowska1982}  I. Sosnowska, T. Neumaier, and E. Steichele, J. Phys. C {\bf 15}, 4835 (1982).
	
	\bibitem{Seidel2009} J. Seidel, L. W. Martin, Q. He, Q. Zhan, Y.-H. Chu, A. Rother, M. E. Hawkridge, P. Maksymovych, P. Yu, M. Gajek, N. Balke, S. V. Kalinin, S. Gemming, F. Wang, G. Catalan, J. F. Scott, N. A. Spaldin, J. Orenstein, and R. Ramesh, Nat. Mat. {\bf 8}, 229 (2009).
		
	\bibitem{Rovillain2010} P. Rovillain, R. de Sousa, Y. Gallais, A. Sacuto, M. A. M\'easson, D. Colson, A. Forget, M. Bibes, A. Barth\'el\'emy, M. Cazayous, Nat. Mat. {\bf 9}, 975 (2010).
		
	\bibitem{Crassous2011} A. Crassous, R. Bernard, S. Fusil, K. Bouzehouane, D. Le Bourdais, S. Enouz-Vedrenne, J. Briatico, M. Bibes, A. Barth\'el\'emy, and J. E.  Villegas, Phys. Rev. Lett. {\bf 107}, 247002 (2011).
		
	\bibitem{Choi2009} T. Choi, S. Lee, Y. J. Choi, V. Kiryukhin, and S.-W. Cheong, Science {\bf 324}, 63 (2009).
		
	\bibitem{Allibe2010} J. Allibe, K. Bougot-Robin, E. Jacquet, I. C. Infante, S. Fusil, C. Carretero, J.-L. Reverchon, B. Marchihac, D. Crete, J.-C. Mage, A. Barthelemy, and M. Bibes, Appl. Phys. Lett. {\bf 96}, 182902 (2010).
		
	\bibitem{Talbayev} D. Talbayev, S. Lee, S.-W. Cheong, and A. J. Taylor, Appl. Phys. Lett. {\bf 93}, 212906 (2008).
		
	\bibitem{Kundys}  B. Kundys, M. Viret, D. Colson, and D. O. Kundys, Nat. Mat. {\bf 9}, 803 (2010).
		
	\bibitem{Lejman} M. Lejman, G. Vaudel, I. C. Infante, P. Gemeiner, V. E. Gusev, B. Dkhil and P. Ruello, Nat. Comm. {\bf 5}, 4301 (2014). 
		
	\bibitem{C8NR05737K} Vila-Fungueiriño, José Manuel and Gómez, Andrés and Antoja-Lleonart, Jordi and Gázquez, Jaume and Magén, César and Noheda, Beatriz and Carretero-Genevrier, Adrián", Nanoscale {\bf 10}, 20155 (2018).
		
	\bibitem{Ruello2012} P. Ruello, T. Pezeril, S. Avanesyan, G. Vaudel, V. Gusev, I. C. Infante, and B. Dkhil, Appl. Phys. Lett. {\bf 100}, 212906 (2012).
		
	\bibitem{Borissenko} E. Borissenko, M. Goffinet, A. Bosak, P. Rovillain, M. Cazayous, D. Colson, P. Ghosez and M. Krisch, J. Phys.: Condens. Matter {\bf 25}, 102201 (2013).
		
		
		
		
		
		
	\bibitem{Rovillain2009} P.Rovillain, M. Cazayous, A. Sacuto, D.Lebeugle, and D.Colson, J. Mag. Mag. Mater. {\bf 21}, 1699 (2009).
		
	\bibitem{Guennou} M. Guennou, H. Dammak, P. Djemia, P. Moch, and M. Pham-Thi, Solid State Sci. {\bf 12}, 298 (2010).
	
	\bibitem{Xu}	F. Xu, L. Belliard, D. Fournier, E. Charron, J.-Y. Duquesne, S. Martin, C. Secouard, and B. Perrin, Thin Solid Films  {\bf 548}, 366 (2013).
	
	\bibitem{Amziane} A. Amziane, L. Belliard, F. Decremps, and B. Perrin, Phys. Rev B {\bf 83}, 014102 (2011).

    \bibitem{Belliard} L. Belliard, A. Huynh, B. Perrin, A. Michel, G. Abadias, and C. Jaouen, Phys. Rev. B {\bf 80}, 155424 (2009).
		
	\bibitem{Born} M. Born, K. Huang, Dynamical Theory of Crystal Lattices. Oxford University Press, New York 1988.
		
	\bibitem{Dreizler} R. M. Dreizler and E. K. U. Gross, Density Functional Theory: An Approach to the Quantum Many-Body Problem, Springer, Berlin (1990).
	
	\bibitem{Hafner2008} J. Hafner, Journal of Computational Chemistry, {\bf 29}, p. 2044-2078 (2008).
		
	\bibitem{Kresse} G. Kresse and J. Furthmuller, Comput. Mater. Sci. {\bf 6}, 15 (1996).
		
	\bibitem{Perdew} J. P. Perdew, A. Ruzsinszky, G. I. Csonka, O. A. Vydrov, G. E. Scuseria, L. A. Constantin, X. Zhou, and K. Burke, Phys. Rev. Lett. {\bf 100}, 136406 (2008).
		
	\bibitem{Shang} S. L. Shang, G. Sheng, Y. Wang, L. Q. Chen, and Z. K. Liu, Phys. Rev. B {\bf 80}, 052102 (2009).
	
	\bibitem{Dudarev1998} S. L. Dudarev, G. A. Botton, S. Y. Savrasov, C. J. Humphreys, and A. P. Sutton, Phys. Rev. B $\bf{57}$, 1505 (1998).
	
	\bibitem{Blochl1994} P. E. Bl\"ochl, O. Jepsen, and O. K. Andersen, Phys. Rev. B. $\bf{49}$, 16223 (1994).
		
	\bibitem{djemia2004} P. Djemia, Y. Roussign\'e, G.F. Dirras and K. M. Jackson, J. Appl. Phys. $\bf{95}$, 2324 (2004).
	
	\bibitem{Djemia1998} P. Djemia, PhD Thesis, Universit\'e Paris 13 (1998).
	
	\end{thebibliography}

\section{References}
	\begin{thebibliography}{1}
		
		\bibitem{JAEKEN2016a} 
		Jan W. Jaeken and Stefaan Cottenier, Computer Physics Communications, {\bf 207}, 445 (2016).
		
		\bibitem{Jaeken2016b} Jan W. Jaeken and Stefaan Cottenier, “Solving the Christoffel equation: Phase and group velocities”, Mendeley Data, V1, (2016) doi: 10.17632/4z25ff88c4.1
		
		\bibitem{Pace1971} N. G. Pace and G. A. Saunders, J. Phys. Chem. Solids {\bf 32}, 1585–1601 (1971).
		
		\bibitem{Ruello2012} P. Ruello, T. Pezeril, S. Avanesyan, G. Vaudel, V. Gusev, I. C. Infante, and B. Dkhil, Appl. Phys. Lett. {\bf 100}, 212906 (2012).
		
		\bibitem{Rivera1997} J.-P. Rivera and H. Schmid, Ferroelectrics,  {\bf 204}, 23-33 (1997)
		
		\bibitem{Kumar2008} A. Kumar, R. C. Rai, N. J. Podraza, S. Denev, M. Ramirez, Y.-H. Chu, L.W. Martin, J. Ihlefeld, T. Heeg, J. Schubert, D. G. Schlom, J. Orenstein, R.
		Ramesh, R. W. Collins, J. L. Musfeldt, and V. Gopalan, Appl. Phys. Lett.
		{\bf 92}, 121915 (2008).
		
		\bibitem{Gu2009} B. Gu, Y. Wang, J. Wang, and W. Ji, Opt. Express {\bf 17}, 10970 (2009).
		
		\bibitem{Djemia2004} P. Djemia, Y. Roussign\'e, G.F. Dirras and K.M. Jackson, J. Appl. Phys. $\bf{95}$, 2324 (2004).
		
		\bibitem{Djemia2001} P. Djemia, F. Ganot, P. Moch, V. Branger and P. Goudeau, J. Appl. Phys.  {\bf 90}, 756–762 (2001).
		
		\bibitem{Djemia1998} P. Djemia, PhD Thesis, Université Paris 13 (1998).
		
	\end{thebibliography}
\end{document}